\documentclass[american,aps,pra,twocolumn]{revtex4-1}
\usepackage[unicode=true,pdfusetitle, bookmarks=true,bookmarksnumbered=false,bookmarksopen=false, breaklinks=false,pdfborder={0 0 0},backref=false,colorlinks=false] {hyperref}
\hypersetup{ colorlinks,linkcolor=myurlcolor,citecolor=myurlcolor,urlcolor=myurlcolor}
\usepackage{graphics,graphicx, amsthm, amsmath, amssymb, times, braket, colortbl, color, bm, soul}
\usepackage[up]{subfigure}
\usepackage{cleveref}
\usepackage[outdir=./]{epstopdf}

\definecolor{myurlcolor}{rgb}{0,0,0.7}

\theoremstyle{plain}

\providecommand{\theoremname}{Theorem}

\newcommand*{\myproofname}{Proof}

\makeatother

\begin{document}

\title{Coherence as witness for quantumness of gravity}

\author{Ahana Ghoshal, Arun Kumar Pati, Ujjwal Sen}

\affiliation{Harish-Chandra Research Institute, HBNI, Chhatnag Road, Jhunsi, Allahabad 211 019, India}

\begin{abstract}
We propose an interferometric set-up that utilizes the concept of quantum coherence to provide quantum signatures of gravity. The gravitational force comes into nontrivial play due to the existence of an extra mass in the set-up that transforms an incoherent state to a coherent state. The implication uses the fact that quantum coherence at a certain site cannot be altered by local actions at a separate site.
The ability to transform an incoherent state to a coherent one in the presence of gravitational field
provides a signature of quantumness of gravity. We also observe that the results remain unaltered in presence of a nontrivial quantity of depolarising noise.
\end{abstract}

\maketitle
\section{Introduction}

Even though quantum theory has been successfully formalized for strong, weak, and electromagnetic fields, there has long been an issue about how to unify quantum theory with gravity.
There are many controversies and curiosities that surround the question of quantum aspects of gravity \cite{Cimento, Diosi, Penrose}. In this respect, one remembers the experiment in 1975 by R. Colella, A. W. Overhauser and S. A. Werner, with a neutron beam, split by an interferometer \cite{Colella}. It was an interference experiment where the phase of the neutron wave function was affected by the gravitational potential. For further experiments in this direction, see \cite{Burgard, Zhou, Zych}. It has been argued that gravity is purely classical in their experiments. For discussions on quantum aspects of gravity and possible ramifications, see \cite{Dewitt, Oriti, Kiefer, Schulz, Casares}.\par
The ubiquitous feature of quantum theory is superposition, and, in particular, the entangled states that it leads to.
%and entanglement. 
Entanglement is a phenomenon that is observed in systems of two or more parts, and is a clear signature of  quantum nature of the system \cite{Horodecki}. In this respect, 
it has been shown that 
``classically mediated gravitational interaction between two gravitationally coupled resonators cannot create entanglement" \cite{Kafri}. Models which support the relativistic semiclassical theory of gravity were given in Ref. \cite{Tilloy}. (See also \cite{D}.) One of the central dogmas of entanglement theory is that 
local quantum operations and classical communication (LOCC) cannot create entanglement \cite{Horodecki}. If any state is unentangled initially, then LOCC can create only a separable state out of it. So, if any unentangled state results in an entangled state after the action of any field, then the field is definitely a quantum entity. Refs. \cite{Bose, Vedral} (see also \cite{Giampaolo}) proposed a thought experiment, which has been 
called the BMV (Bose \emph{et al.} - Marletto - Vedral) effect, in which two massive particles were sent through two interferometers. At the end of the interferometry, they found that there is entanglement between the path degrees of freedom of the two particles, while initially the particles were unentangled. From this result, they concluded that the gravitational interaction between the two particles must be of quantum nature, as it can create entanglement. A simple  analysis of these aspects is found in Ref. \cite{Rovelli}. See also 
\cite{questions-answers}.
\par
Just like entanglement, quantum coherence is also an exclusively quantum phenomenon. Both quantum coherence and quantum entanglement arise from superposition. 
%See Ref. \cite{Aberg,Cramer} for a 
A formal definition and further ramifications of the concepts around quantum coherence can be found in 
Refs.  \cite{Aberg,Cramer}.
See \cite{Carlesso} in this regard. The concept of quantum coherence should not be confused with the ``coherent states" in quantum optical systems \cite{Scully}. \par  
In this paper, we show that gravitational interaction between two particles can create a quantum coherent state with respect to some basis, while in absence of one of the particles, there occurs an incoherent state with respect to the same basis. We quantify the amount of coherence created by using two distance-based measures of the same. Thus, we argue that the ability to transform an incoherent state into a 
coherent one in the presence of a gravitational field is a signature of quantumness of gravity.
%\section{
\par
\section{A massive particle sent through a beam splitter}
\label{sec_1}
%\par
%In this section, 
We begin by considering the gravitational interaction acting between two components of the same particle.
\begin{figure}
\includegraphics[width=9cm,height=7cm]{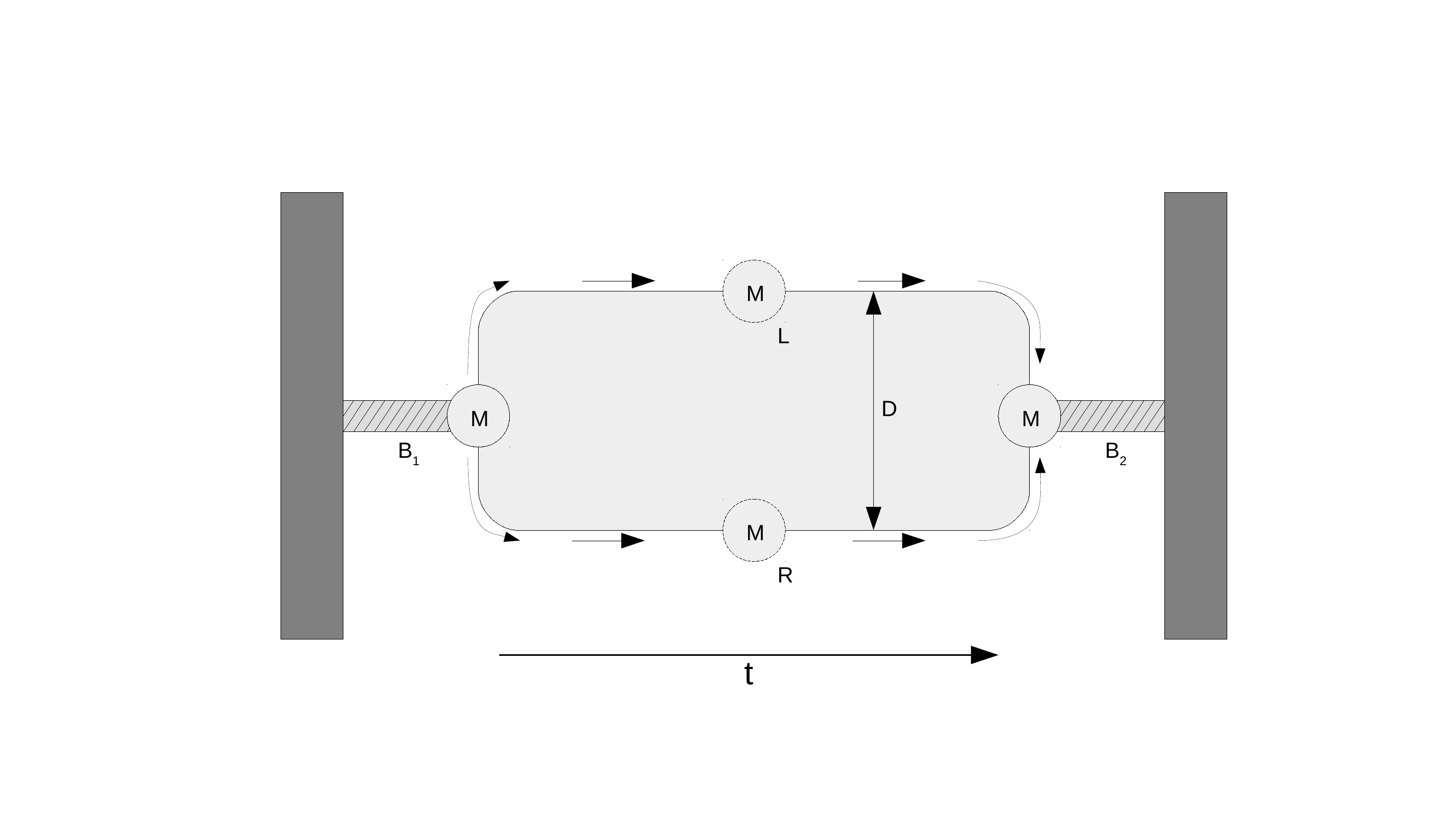}
\caption{The arrangement of the experiment in which a particle of mass $M$ is transmitted through the beam splitter $B_1$. The distance between the two arms of $B_1$ is $D$ in which the split parts of $M$ are traveling. The two components of the mass are gathered by the beam splitter $B_2$. Here, the exact form of $B_1$ and $B_2$ will depend on the type of particles used. If they are neutrons, then $B_1$ and $B_2$ will be as in \cite{Colella}.} 
\label{fig1}
\end{figure}
Consider a mass $M$ moving through a beam splitter, by which the particle is split in two spatially separated components $|L\rangle$ and $|R\rangle$ as shown in Fig. \ref{fig1}. The distance of the centres of the two components is $D$ and $|L\rangle$ and $|R\rangle$ are orthogonal states i.e,. $\langle L|R\rangle=0$. We can assume that each part is a localized Gaussian wave packet with width, $\Delta x \ll D$, as in \cite{Bose}. So, we can assume that the entire system is in the state,
\begin{equation}
|\psi(t=0)\rangle=\frac{1}{\sqrt{2}}(|L\rangle+|R\rangle)\otimes|g\rangle,
\label{equ_1}
\end{equation}  
where $|g\rangle$ is the quantum state of the gravitational field on each component of the particle due to the other. The energy of each component will be the gravitational potential energy due to the existence of the other.
Therefore, energy of each component is
\begin{equation}
E_L=E_R=E
=\frac{GM^2}{D}.
\end{equation}
Since physical spacetime geometry can be in superposition of macroscopically distinct configurations \cite{Rovelli}, at time $t=\tau$, the state will be
\begin{equation}
|\psi(t=\tau)\rangle=\frac{1}{\sqrt{2}}(e^{-\frac{iE_L\tau}{\hbar}}|L\rangle|g_L\rangle+e^{-\frac{iE_R\tau}{\hbar}}|R\rangle|g_R\rangle).
\end{equation}
As the masses of the two components of the particle are considered to be same, we have $|g_L\rangle =|g_R\rangle $.
The state at time $t=\tau$ is
\begin{equation}
|\psi(t=\tau)\rangle=\frac{1}{\sqrt{2}}e^{-\frac{iE\tau}{\hbar}}(|L\rangle|g_L\rangle+|R\rangle|g_R\rangle),
\label{equ_4}
\end{equation}
where we can now neglect the overall phase. The two components of the particle are brought back together at $B_2$. Hence, the wave function, then, is,
\begin{equation}
|\psi(t=\tau)\rangle=\frac{1}{\sqrt{2}}(|L\rangle+|R\rangle)\otimes|g\rangle.
\end{equation} 
If the gravitational field is traced out, the particle state at time $t=\tau$ will be
\begin{equation}
|\psi_1\rangle=\frac{1}{\sqrt{2}}(|L\rangle+|R\rangle).
\end{equation}
So, the corresponding density matrix is
\begin{equation}
\rho_1=|\psi_1\rangle\langle\psi_1|=\frac{1}{2}(|L\rangle+|R\rangle)(\langle L|+\langle R|).
\end{equation}
Now, we will compute the coherence of this state with respect to the basis $\lbrace \frac{(|L\rangle\pm|R\rangle)}{\sqrt{2}}\rbrace $.
Here, we will consider two measures of coherence defined in Ref. \cite{Cramer}.\par
Quantum coherence of the state of a quantum system is the existence of off-diagonal terms in the density matrix of the state. An understanding of its presence was known since the beginnings of quantum mechanics, and was known to the reason for several phenomena including interference. However, the modern theory of quantum coherence is relatively new \cite{Aberg, Cramer}, and has also partially fed the interesting stream of research on resource theories. 
It is clear that whether a quantum state possesses coherence depends on the choice of basis. There are many ways in which one can quantify coherence, and we choose two such instances.\\
$\bullet$ \textbf{Relative entropy of coherence:} Let $\hat{\rho}$ be a density matrix written in some basis. The relative entropy of coherence of $\hat{\rho}$ in that basis is 
\begin{equation}
C_{rel.ent.}(\hat{\rho})=S(\hat{\rho}_{diag})-S(\hat{\rho}),
\end{equation}
where $S(\cdot)$ is the von Neumann entropy of its argument and $\hat{\rho}_{diag}$ is the state obtained from $\hat{\rho}$ by removing the off-diagonal elements.\par
Here, in our case $\rho_1$ is diagonal in the basis $\lbrace \frac{(|L\rangle\pm|R\rangle)}{\sqrt{2}}\rbrace $. So, $S(\rho_{1_{diag}})$ and $S(\rho_1)$ will give the same value, and we get $C_{rel.ent.}(\rho_1)=0$.\\
$\bullet$ \textbf{$\mathbf{l_1}$-norm coherence:} The $l_1$-norm coherence of $\hat{\rho}$ in a given basis is defined as
\begin{equation}
C_{l_1}(\hat{\rho})=\sum_{\substack{{i,j}\\
					i\neq j}}
					|\rho_{i,j}|,
\end{equation}  
i.e, it is the sum of the moduli of all nonzero non-diagonal elements of $\hat{\rho}$, where $\hat{\rho}$ is expressed in the given basis.
In our case, $\rho_1$ is diagonal in the basis $\lbrace \frac{(|L\rangle\pm|R\rangle)}{\sqrt{2}}\rbrace $. So, $C_{l_1}(\rho_1)=0$.\par 
We therefore see that both the measures of coherence indicate that we have obtained an incoherent state with respect to the basis $\lbrace \frac{(|L\rangle\pm|R\rangle)}{\sqrt{2}}\rbrace $. 
%We will next see what happens if an extra mass is present in this above experiment. 
We next consider the implications of noise in the set-up, after which we try to find the outcome of having an extra mass in the above thought experiment. 
\par
\subsection{Response to noise}
\label{Noise}
It is plausible that the system will be affected by noise, resulting in the initial state to be a mixed state instead of a pure one. Assuming that the noise is completely depolarized, the initial state, instead of being given by Eq. (\ref{equ_1}), will be given by
\begin{equation}
\rho(t=0)= \Big[p\frac{(|L\rangle+|R\rangle)}{\sqrt{2}}\frac{(\langle L|+\langle R|)}{\sqrt{2}}+(1-p)\frac{I_2}{2}\Big] \otimes |g\rangle\langle g|, 
\end{equation}
where $0\leq p\leq 1$ and where $I_2$ can be written as $I_2=|L\rangle\langle L|+|R\rangle\langle R|$. The corresponding state in place of the one in Eq. (\ref{equ_4})  at time $t=\tau$, is
\begin{equation}
\rho(t=\tau)=\Big[\frac{1}{2}(|L\rangle\langle L|+|R\rangle\langle R|)+\frac{p}{2}(|L\rangle\langle R|+|R\rangle\langle L|)\Big]\otimes |g\rangle\langle g|.
\end{equation} 
After tracing out the gravitational field, the particle state at time $t=\tau$ is
\begin{equation}
\rho_{1_{Noisy}}(t=\tau)=\Big[\frac{1}{2}(|L\rangle\langle L|+|R\rangle\langle R|)+\frac{p}{2}(|L\rangle\langle R|+|R\rangle\langle L|)\Big],
\end{equation} 
which is again diagonal in the basis $\lbrace \frac{(|L\rangle\pm|R\rangle)}{\sqrt{2}}\rbrace $. So, both relative entropy of coherence %$C_{rel.ent.}(\tilde{\rho}_1)=0$ 
and $\mathbf{l_1}$-norm coherence 
%$C_{l_1}(\tilde{\rho}_1)=0$. 
 vanish even in the presence of noise. 
 We therefore 
 %So, we can 
 see that the result with an initial noisy state, in the set-up considered in Fig. \ref{fig1} and with the noise model considered, is the same as with the initial pure state.  
\par
\section{A massive particle sent through a beam splitter with an extra mass running parallel to the split particle}
\label{Sec_2}
\begin{figure}
\includegraphics[width=9cm,height=7cm]{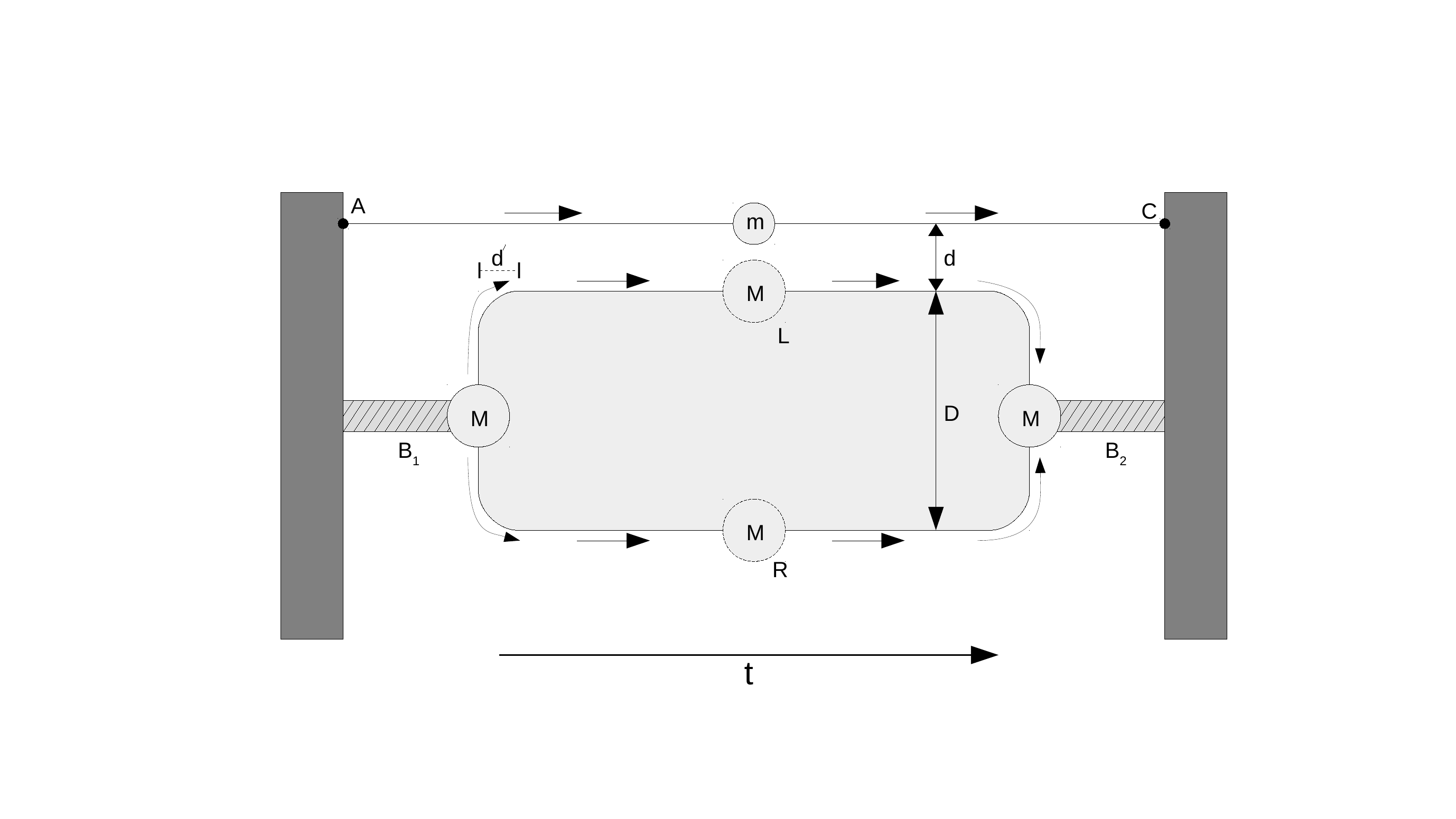}
\caption{The set-up is just as in Fig. \ref{fig1}, except that an extra mass $m$ is running parallel to the split components of mass $M$ through a channel $AC$. The masses \(m\) and \(M\) are moving together on different channels and at the same speeds. $d$ is the distance between the centre of mass of $m$ and the $L$ component of $M$. $d$ is very small compared to $D$. $d'$ is the length of nonparallel part of the arms of the beam splitter $B_1$. $d'$ is assumed to be so small that we can neglect this length. All other considerations are the same as in Fig. \ref{fig1}.} 
\label{fig2}
\end{figure}

%In this section, 
We now consider a set-up which has all the components of the preceding one, but has an additional feature. In the current set-up, a particle of mass $m$ moves in  parallel to the split components of the particle of mass $M$. The experimental set-up is described in Fig. \ref{fig2}.
The initial state of the entire system is
\begin{equation}
|\tilde{\psi}(t=0)\rangle=|m\rangle\otimes\frac{1}{\sqrt{2}}(|L\rangle+|R\rangle)\otimes|\tilde{g}\rangle.
\end{equation}
Here, $|\tilde{g}\rangle$ is the quantum state of the gravitational field due to the mass $m$ on the components of mass $M$. As the distances of the two components ($L$ and $R$) of mass $M$ are different from the mass $m$, carried by $AC$, the quantum states of the gravitational field are different in the $L$ and $R$ channels. 
Now, the gravitational potential energies on each component of $M$, due to the existence of the extra mass, are
\begin{equation}
E_L=\frac{GMm}{d} \qquad \textrm{and} \qquad E_R=\frac{GMm}{d+D}.
\end{equation}
Here, we have ignored the gravitational potential energy on one component of the particle due to the other. As in the previous case, it will introduce an overall phase, which we can neglect.
Since the metric in different branches of the interferometer represent distinct spacetimes, 
 at time $t=\tau$, the state will be
\begin{equation}
|\tilde{\psi}(t=\tau)\rangle=\frac{1}{\sqrt{2}}(e^{-i\phi_L}|m\rangle|L\rangle|\tilde{g}_L\rangle+e^{-i\phi_R}|m\rangle|R\rangle|\tilde{g}_R\rangle),
\end{equation}
where
\begin{eqnarray*}
&&\phi_L=\frac{GMm\tau}{\hbar d} \qquad \textrm{and} \qquad \phi_R=\frac{GMm\tau}{\hbar (d+D)},\\
\textrm{with} \qquad &&\Delta\phi=\phi_L-\phi_R=\frac{GMm\tau D}{\hbar d(d+D)}.
\end{eqnarray*}
%with $\Delta\phi=\phi_L-\phi_R=\frac{GMm\tau\Delta x}{d(d+\Delta x)}$.
The state  of the combined system is
\begin{equation}
|\tilde{\psi}(t=\tau)\rangle=\frac{1}{\sqrt{2}}e^{-i\phi_L}(|m\rangle|L\rangle|\tilde{g}_L\rangle+e^{i\Delta\phi}|m\rangle|R\rangle|\tilde{g}_R\rangle).
\label{equ_16}
\end{equation}
The two components of the particle of mass $M$ are brought back together. Then, the state will be
\begin{equation}
|\tilde{\psi}(t=\tau)\rangle=\frac{1}{\sqrt{2}}(|m\rangle|L\rangle+e^{i\Delta\phi}|m\rangle|R\rangle)\otimes |\tilde{g}\rangle.
\end{equation}
As in the previous case, we trace out the gravitational field, to get
\begin{equation}
|\tilde{\psi}_1\rangle=\frac{1}{\sqrt{2}}(|m\rangle|L\rangle+e^{i\Delta\phi}|m\rangle|R\rangle).
\end{equation}
Then, tracing out the degrees of freedom of the particle carried by the channel $AC$, we get the state of the particle of mass $M$, as
\begin{equation}
|\tilde{\psi}_2\rangle=\frac{1}{\sqrt{2}}(|L\rangle+e^{i\Delta\phi}|R\rangle).
\label{equ_19}
\end{equation}
The corresponding density matrix is
\begin{equation}
\tilde{\rho}_2=|\tilde{\psi}_2\rangle\langle\tilde{\psi}_2|=\frac{1}{2}(|L\rangle+e^{i\Delta\phi}|R\rangle)(\langle L|+e^{-i\Delta\phi}\langle R|).
\end{equation}
If we choose the $\lbrace \frac{(|L\rangle\pm|R\rangle)}{\sqrt{2}}\rbrace$ basis, then the state $\tilde{\rho}_2$ can be expressed as
\begin{equation}
        \tilde{\rho}_2=
  \left( {\begin{array}{cc}
        \rho_{11}  & \rho_{12} \\
        \rho_{21} & \rho_{22}
        \end{array} } \right).
\end{equation}
The four elements of the density matrix are
\begin{eqnarray*}
\rho_{11}&=&\cos^{2}(\frac{\Delta\phi}{2}),\\
\rho_{12}&=&\frac{i}{2}\sin (\Delta\phi),\\
\rho_{21}&=&-\frac{i}{2}\sin (\Delta\phi),\\
\rho_{22}&=&\sin^{2}(\frac{\Delta\phi}{2}).
\end{eqnarray*}
So, $\tilde{\rho}_2$ is a non-diagonal matrix in the $\lbrace \frac{(|L\rangle\pm|R\rangle)}{\sqrt{2}}\rbrace$ basis. Hence, $\tilde{\rho}_2$ is coherent in this basis.\\
$\bullet$ \textbf{Relative entropy of coherence:} The relative entropy of coherence of $\tilde{\rho}_2$ is
\begin{eqnarray}
\nonumber
C_{rel.ent.}(\tilde{\rho}_2)&=&S(\tilde{\rho}_{2_{diag}})-S(\tilde{\rho}_2)\\
\nonumber
&=&-(1+\cos(\Delta\phi))\log_2[\cos(\frac{\Delta\phi}{2})]\\
&&-(1-\cos(\Delta\phi))\log_2[\sin(\frac{\Delta\phi}{2})].
\label{equ_22}
\end{eqnarray}
$\bullet$ \textbf{$\mathbf{l_1}$-norm coherence:} The $l_1$-norm coherence of $\tilde{\rho}_2$ is
\begin{eqnarray}
\nonumber
C_{l_1}(\tilde{\rho}_2)&=&|\frac{i}{2}\sin(\Delta\phi)|+|-\frac{i}{2}\sin(\Delta\phi)|\\
&=&|\sin(\Delta\phi)|.
\label{equ_23}
\end{eqnarray} 
So, according to both measures of coherence, we get a coherent state with respect to the $\lbrace \frac{(|L\rangle\pm|R\rangle)}{\sqrt{2}}\rbrace$ basis.
Therefore, the presence of gravity can transform an incoherent state to a state having nonzero coherence.
%\textcolor{red}{
If the gravitational field due to \(m\) acts classically on \(M\), 
with the \(M\) having already split at the beam splitter \(B_1\), then the two arms of the interferometer 
can only be acted on by a multiple of the identity.
In other words, the relevant Hamiltonian in the classical case 
%of the system 
acts on $|\tilde{\psi}(t=0)\rangle$ as an identity operator, and at time $t=\tau$, the initial state remains the same up to an overall phase, and no relative phase can be introduced. See the appendix for 
more details. So, there is no possibility of increasing the coherence of the two-level system formed by 
\(M\) by a classical field due to \(m\).
%}
\par
\subsection{In presence of noise}
As in 
%Sec. \ref{Noise}, 
the noisy scenario considered for the set-up in Fig. \ref{fig1},
we now consider an initial noisy state of the form,
\begin{eqnarray}
\nonumber
\tilde{\rho}(t=0)=|m\rangle\langle m|&\otimes &\Big[p\frac{(|L\rangle+|R\rangle)}{\sqrt{2}}\frac{(\langle L|+\langle R|)}{\sqrt{2}}\\&&+(1-p)\frac{I_2}{2}\Big] \otimes |\tilde{g}\rangle\langle \tilde{g}|.
\end{eqnarray}
The corresponding state in place of the one in Eq. (\ref{equ_16}), i.e., the state of the composite system at time $t=\tau$, is
\begin{eqnarray}
\nonumber
\tilde{\rho}(t=\tau)&=&|m\rangle\langle m|\otimes \Big[\frac{1}{2}\big(|L\rangle\langle L|\otimes |\tilde{g}_L\rangle\langle \tilde{g}_L|\\
\nonumber
&+&|R\rangle\langle R|\otimes |\tilde{g}_R\rangle\langle \tilde{g}_R|+\frac{p}{2}e^{-i\Delta \phi} |L\rangle\langle R| \otimes |\tilde{g}_L\rangle\langle \tilde{g}_R|\\
&+&\frac{p}{2}e^{i\Delta \phi} |R\rangle\langle L| \otimes |\tilde{g}_R\rangle\langle \tilde{g}_L| \Big].
\end{eqnarray}
So, in the noisy scenario, in place of Eq. (\ref{equ_19}) in the noiseless case, we have
\begin{eqnarray}
\nonumber
\tilde{\rho}_{2_{Noisy}}&=&\Big[\frac{1}{2}\big(|L\rangle\langle L|+|R\rangle\langle R|\big)\\&+&\frac{p}{2}e^{-i\Delta \phi} |L\rangle\langle R|+\frac{p}{2}e^{i\Delta \phi} |R\rangle\langle L|\Big].
\end{eqnarray}
$\tilde{\rho}_{2_{Noisy}}$ is non-diagonal in the $\lbrace \frac{(|L\rangle\pm|R\rangle)}{\sqrt{2}}\rbrace$ basis. In that basis, it has the matrix form,
\begin{equation}
        \tilde{\rho}_{2_{Noisy}}=
  \left( {\begin{array}{cc}
        \frac{1}{2}+\frac{p}{2}\cos(\Delta\phi)  & \frac{p}{2}i\sin(\Delta\phi) \\
        -\frac{p}{2}i\sin(\Delta\phi) & \frac{1}{2}-\frac{p}{2}\cos(\Delta\phi)
        \end{array} } \right).
\end{equation}
The relative entropy of coherence in this case is
\begin{eqnarray}
\nonumber
C_{rel.ent.}&&(\tilde{\rho}_{2_{Noisy}})=\frac{1-p}{2}\log_2(1-p)+\frac{1+p}{2}\log_2(1+p)\\
\nonumber
&&-(\frac{1}{2}+\frac{p}{2}\cos(\Delta\phi))\log_2[1+p\cos(\Delta\phi)]\\
&&-(\frac{1}{2}-\frac{p}{2}\cos(\Delta\phi))\log_2[1-p\cos(\Delta\phi)].
\label{equ_28}
\end{eqnarray}
And the $\mathbf{l_1}$-norm coherence is given by
\begin{equation}
C_{l_1}(\tilde{\rho}_{2_Noise})=p|\sin(\Delta\phi)|.
\label{equ_29}
\end{equation} 
For both the measures the coherence, the output state in the noisy case is less coherent than that in the noiseless scenario, for all $0\leq p <1$. Compare Eqs. (\ref{equ_28}) and (\ref{equ_29}) with (\ref{equ_22}) and (\ref{equ_23}). However for any $p\neq 0$, the coherence, according to both the measures, in the state $\tilde{\rho}_{2_{Noisy}}$ is strictly greater than that in $\rho_{1_{Noisy}}$. 
\par
Let us add here the following observation.
The relative phase, \(\Delta \phi\), is given by 
\begin{equation}
\Delta \phi = \frac{GMm \tau D}{\hbar d (d+D)}.
\end{equation} 
This can be re-written as 
\begin{equation}
\Delta \phi = \frac{\alpha Mm }{m_P^2},
\end{equation}
where \(\alpha = (\tau D c)/(d(d+D))\) is a dimensionless parameter and \(m_P\) is the Planck mass given by 
\(m_P^2 = \hbar c/G\). With suitable choice of the parameters, it is possible to make \(\Delta \phi = \pi/2\), whereby we have a maximal coherence of unity in the output, in the noiseless case. 
\par
Another comment is in order here. Quantum coherence is basis-dependent quantity. Therefore, the 
numbers obtained are altered if we change the basis. In particular, if we choose the basis \(\{|L\rangle, |R\rangle\}\), then the outputs in both the situations considered (i.e., in Figs. \ref{fig1} and \ref{fig2}) will 
have nonzero quantum coherence. However, the particular basis that we have chosen provides a stark 
zero versus nonzero constrast between the two situations, even in the noisy case.
 
\section{Conclusion}
In this paper, we have shown that presence
of the gravitational force due to a mass in the neighborhood
of a particle, split by a beam splitter, can create a coherent
state with respect to some basis, while the absence of the mass
will give an incoherent state with respect to the same basis. It
is to be noted that a local action at a certain site cannot create
coherence at a separated site. Therefore, the increase of
coherence must have been created due to a gravitational interaction of
quantum origin
between the two particles. Our result is based on
two assumptions, viz. (i) in the perturbative regime, Newtonian
theory holds, and (ii) superposition of distinct spacetimes
is possible. We have also demonstrated that any nontrivial
amount of depolarizing noise keeps the results unchanged
qualitatively, although the quantities do alter. In the appendix, we have shown that our result cannot be explained by assuming
that gravitational interaction is classical, and if one can explain the
creation of coherence by classical nature of gravity, then the same
argument applies to creation of entanglement between two distinct masses.
Since controlling
entanglement is a difficult task, we hope that our scheme
can be tested with relative ease, in comparison to those in
Refs. \cite{Bose, Vedral}.

%In this paper, we have seen that the presence of the gravitational force due to a mass in the neighborhood of a particle, split by a beam splitter, can create a coherent state with respect to some basis, while the absence of the mass will give an incoherent state with respect to the same basis. It is to be noted that a local action at a certain site cannot create coherence at a separated site. Therefore, the increase of coherence must have been created due to the gravitational interaction between the two particles.
%Our result is based on two assumptions, viz. (i) in the perturbative regime, Newtonian theory holds, and (ii) superposition of distinct spacetimes is possible.
%We have also demonstrated that any nontrivial amount of depolarising noise keeps the results unchanged qualitatively, although the quantities do alter.
%\textcolor{red}{We also found that the presence of noise in an initial pure state does not make an incoherent state coherent, but it reduces the coherence of a coherent state.}
%Since controlling entanglement is a difficult task, we hope that our scheme can be tested with relative ease, in comparison to those in Refs. \cite{Bose, Vedral}.

We acknowledge useful discussions with Sougato Bose at the Raman Research Institute, Bengaluru, India.

\begin{center}
\textbf{Appendix}
\end{center}

\emph{\textbf{The classical approach:}} In this 
%Supplementary Material, 
appendix
we wish to discuss a few aspects about the effect of gravitational fields, if it were classical, on coherence in the system on which the field is acting. 

%There may arise some confusion, because sometimes classical fields can change the phase, hence can increase coherence. Let us discuss about the classical treatment. 
To understand the classical case,
%approach of gravitational field 
let us first consider a set-up from Ref. \cite{Kafri}, where they have taken two masses, $m_1$ and $m_2$, suspended freely so that their motion can be approximated to a harmonic oscillator moving along the $x$-axis. See Fig. 1 in \cite{Kafri}. The displacement of mass $m_j$ is denoted by $x_j$. The interaction potential energy between the masses is given by
\begin{equation}
V(x_1,x_2)=V_0-\frac{Gm_1m_2}{d^2}(x_1-x_2)-\frac{Gm_1m_2}{d^3}(x_1-x_2)^2,
\end{equation}
where $V(x_1,x_2)$ is expanded to second order in the relative displacement, and where $d$ is the equilibrium distance between the two masses. The quantum mechanical Hamiltonian is
\begin{equation}
H=H_0+K\hat{x}_1\hat{x}_2,
\end{equation} 
where
\begin{equation}
H_0=\sum_{j=1}^{2}\frac{\hat{p}_j^2}{2m_j}+\frac{m_j(\omega_j^2-\frac{K}{m_j})}{2}\hat{x}_j^2,
\end{equation}
with $\omega_j$ being the frequency of oscillation of the $j^{th}$ oscillator and $K=\frac{2Gm_1m_2}{d^3}$.
The quantum interaction term of the form $\hat{x}_1\hat{x}_2$ is assumed to be mediated by a classical channel, with the derivation using methods from quantum stochastic control theory, in Ref. \cite{Kafri}. The theory tells that ``the gravitational centre of mass co-ordinate, $x_j$, of each particle is continuously measured, and a classical stochastic measurement record, $K_j$, carrying this information, acts reciprocally as a classical control force on the other mass.''

According to this theory, the Hamiltonian, in the case when the gravitational fields act classically, in the situation where an extra mass is moving through a channel parallel to the split mass, in the set-up in Fig. 2 of our manuscript, is
\begin{equation}
H_{cl}=\sum_{\substack{j}} K^{\prime}_j(t)x_j, 
\end{equation}
where \(j\) runs over 1 and 2, with \(j=1\) denoting the position variable of the mass \(m\), while \(j=2\) that of the split mass \(M\). 
Here we take $\int_{0}^{t} K^{\prime}_j(t) dt=K_j(t)$. So the unitary operator for time evolution is $e^{\frac{i}{\hbar}\int_{0}^{t}\sum_{\substack{j}} K^{\prime}_j(t)x_j dt}$.

The initial (unnormalized) state in position basis can, for example, be assumed to be 
\begin{equation}
\langle x,x^{\prime}|\psi(t=0)\rangle _{cl}=e^\frac{-(x-\overline{x}_1)^2}{2\sigma}\otimes\left[e^\frac{-(x^{\prime}-\overline{x}_2)^2}{2\sigma}+e^\frac{-(x^{\prime}-\overline{x}_3)^2}{2\sigma}\right],  
\end{equation} 
with a suitable standard deviation \(\sigma\) of the Gaussian waveforms, and 
where $\overline{x}_1$, $\overline{x}_2$ and $\overline{x}_3$ are the equilibrium positions of mass $m$, and the $|L\rangle$ and $|R\rangle$ components of mass $M$, respectively. The exact details of the waveforms, however, are not relevant for 
the conclusion.
So, the final state at time $t=\tau$ is
\begin{equation}
\begin{aligned}
\langle x,x^{\prime}|\psi(t=\tau)\rangle _{cl}={}&e^{{-\frac{i}{\hbar}}\sum_{\substack{j}} K_j(t)x_j}\langle x,x^{\prime}|\psi(t=0)\rangle _{cl}\\
={}&e^{-\frac{i}{\hbar}K_1(t)x}e^{-\frac{i}{\hbar}K_2(t)x^{\prime}}\langle x,x^{\prime}|\psi(t=0)\rangle _{cl}.
\end{aligned} 
\end{equation} 
Hence the time evolution of the initial state introduces only a global phase, which does not change the coherence of the state of the split mass \(M\). 
Similarly, if we take the set-up in the BMV effect \cite{Bose, Vedral}, with two split masses, the time evolution of the initial state by a Hamiltonian, in the case when the gravitational fields act classically, will also give a global phase, and an initial unentangled state will remain unentangled. This proves that if we treat the action of a gravitational field as a classical channel, it cannot create entanglement between the sites at the ends of the channel, and cannot create coherence at a distant site at the other end of the channel.\par
%%%%%%%%%%%%%%%%%%%%%%%%%%%%%%%%%%%%%%%%%%%%%%%%%%%%%%%%%%%%%%%%%%%%%%%%%%
%Now, we consider a different approach and argue whether this approach is quantum or classical. For the system of our paper if we think to deal with the system classically, then the interaction Hamiltonian should be,
We note here that it is erroneous to assume that in the situation considered in Fig. 2 of the manuscript, the Hamiltonian of the split mass \(M\), on which the gravitational field due to the mass \(m\) is acting classically, is given by
\begin{equation}
\label{harkin-chot}
H=E_L|L\rangle\langle L|+E_R|R\rangle\langle R|.  
\end{equation}
This Hamiltonian, after acting on an initial state, almost surely introduces coherence. Consider, however, the set-up of the two split masses as in the BMV effect \cite{Bose, Vedral}. In this case, the corresponding Hamiltonian of the split mass \(M\), on which the gravitational field due to the split mass \(m\) is acting classically, would then be given by
%
%Now, we have to find whether this type of Hamiltonian can be said to be classical or not. To observe this let us take into account the BMV effect where two masses are split by two beam splitter kept at a distance $d$. The distance between the two arms of the beam splitters are $\Delta x_1$ and $\Delta x_2$. See Fig. 1 in \cite{Bose}. In classical treatment the interaction Hamiltonian for that case should be,
\begin{eqnarray}
\nonumber
H_1&=&|R_2\rangle\langle R_2|\otimes(E_1|R_1\rangle\langle R_1|+E_2|L_1\rangle\langle L_1|)\\
&+&|L_2\rangle\langle L_2|\otimes(E_1^\prime |R_1\rangle\langle R_1|+E_2^\prime |L_1\rangle\langle L_1|). 
\end{eqnarray}  
The two masses are split by two beam splitters kept at a distance $d$. The distance between the two arms of the beam splitters are $\Delta x_1$ and $\Delta x_2$.
%See Fig. 1 in \cite{Bose}.
Choosing $E_1=1$, $E_2=1$,$E_1^\prime=1$ and $E_2^\prime=-1$, the Hamiltonian, in dimensionless units, is
\begin{eqnarray}
\nonumber
H_1&=&|R_2\rangle\langle R_2|\otimes(|R_1\rangle\langle R_1|+|L_1\rangle\langle L_1|)\\
&+&|L_2\rangle\langle L_2|\otimes(|R_1\rangle\langle R_1|-|L_1\rangle\langle L_1|). 
\label{quarantine}
\end{eqnarray}
This Hamiltonian, however, can create an entangled state from an unentangled one, as can be seen, for example, by considering the 
 initial state of the two split masses system as
\begin{equation}
|\psi(t=0)\rangle _{12}=\frac{1}{\sqrt{2}}(|L_1\rangle +|R_1\rangle)\frac{1}{\sqrt{2}}(|L_2\rangle +|R_2\rangle),
\end{equation}
as shown in Ref. \cite{Bose, Vedral}.
%At time $t=\tau$, the state will be,
%\begin{multline}
%|\psi(t=\tau)\rangle _{12}=(e^{-iE_2\tau} |R_2\rangle |L_1\rangle +e^{-iE_1\tau} |R_2\rangle |R_1\rangle\\
%+e^{-iE_1^\prime\tau} |L_2\rangle |R_1\rangle+e^{-iE_2^\prime\tau} |L_2\rangle |L_1\rangle)\\
%=(e^{-i\tau} |R_2\rangle |L_1\rangle +e^{-i\tau} |R_2\rangle |R_1\rangle +e^{-i\tau} |L_2\rangle |R_1\rangle +e^{i\tau} |L_2\rangle |L_1\rangle)\\
%=e^{-i\tau} (|R_2\rangle |L_1\rangle +|R_2\rangle |R_1\rangle +|L_2\rangle |R_1\rangle +e^{2i\tau} |L_2\rangle |L_1\rangle).
%\end{multline}
%For some specific values of $\tau$ we can get $e^{2i\tau}=-1$. So,
%\begin{eqnarray}
%\nonumber
%|\psi(t=\tau)\rangle _{12}&=&e^{-i\tau}(|R_2\rangle |L_1\rangle +|R_2\rangle |R_1\rangle \\
%&+&|L_2\rangle |R_1\rangle -|L_2\rangle |L_1\rangle)
%\end{eqnarray}
%The final state is entangled as Bose et al. got in their paper \cite{Bose}. 
Therefore,  the action of an unsplit \(m\) on a split \(M\) cannot be described by the Hamiltonian in Eq. (\ref{harkin-chot}), just like the action of a split \(m\) on a split \(M\) cannot be described by the Hamiltonian in Eq. (\ref{quarantine}),  if the gravitational field acts classically.
%So, this type of Hamiltonian can create entanglement. Hence, this Hamiltonian cannot be said to be a classical one. 

\end{document}